\documentclass[a4paper,11pt]{article}
\usepackage{amssymb}
\usepackage{latexsym,bm}

\def \be {\begin{equation}}
\def \ee {\end{equation}}
\def \bea {\begin{eqnarray}}
\def \eea {\end{eqnarray}}
\def \nn {\nonumber}

\def \a {\alpha}
\def \b {\beta}

\def \G {\Gamma}
\def \d {\delta}

\def \m {\mu}
\def \n {\nu}
\def \k {\kappa}

\def \s {\sigma}
\def \r {\rho}
\def \o {\omega}

\def \th {\theta}
\def \Th {\Theta}

\def \t {\tau}
\def \dag {\dagger}
\def \p {\partial}

\def\bd{\begin{document}}
\def\ed{\end{document}}
\def\nn{\nonumber}
\def\bea{\begin{eqnarray}}
\def\eea{\end{eqnarray}}
\let\bm=\bibitem
\let\la=\label

\def\N{{\cal N}}
\def\sst{\scriptscriptstyle}
\def\thetabar{\bar\theta}
\def\Tr{{\rm Tr}}
\def\one{\mbox{1 \kern-.59em {\rm l}}}

\def\a{\alpha}      \def\da{{\dot\alpha}}
\def\b{\beta}       \def\db{{\dot\beta}}
\def\c{\gamma}  \def\C{\Gamma}  \def\cdt{\dot\gamma}
\def\d{\delta}  \def\D{\Delta}  \def\ddt{\dot\delta}
\def\e{\epsilon}        \def\vare{\varepsilon}
\def\f{\phi}    \def\F{\Phi}    \def\vvf{\f}
\def\h{\eta}
\def\k{\kappa}
\def\l{\lambda} \def\L{\Lambda}
\def\m{\mu} \def\n{\nu}
\def\o{\omega}
\def\P{\Pi}
\def\r{\rho}
\def\s{\sigma}  \def\S{\Sigma}
\def\t{\tau}
\def\th{\theta} \def\Th{\Theta} \def\vth{\vartheta}
\def\X{\Xeta}
\def\z{\zeta}
\def\w{\wedge}
\def\u{\underline}
\def\hs{\hspace}
\def\lt{\left}

\def\cA{{\cal A}} \def\cB{{\cal B}} \def\cC{{\cal C}}
\def\cD{{\cal D}} \def\cE{{\cal E}} \def\cF{{\cal F}}
\def\cG{{\cal G}} \def\cH{{\cal H}} \def\cI{{\cal I}}
\def\cJ{{\cal J}} \def\cK{{\cal K}} \def\cL{{\cal L}}
\def\cM{{\cal M}} \def\cN{{\cal N}} \def\cO{{\cal O}}
\def\cP{{\cal P}} \def\cQ{{\cal Q}} \def\cR{{\cal R}}
\def\cS{{\cal S}} \def\cT{{\cal T}} \def\cU{{\cal U}}
\def\cV{{\cal V}} \def\cW{{\cal W}} \def\cX{{\cal X}}
\def\cY{{\cal Y}} \def\cZ{{\cal Z}}

\def\ua{\underline{\alpha}} \def\ubb{\underline{\beta}}
\def\ug{\underline{\gamma}}
\def\ub{\underline{\phantom{\alpha}}\!\!\!\beta}
\def\uc{\underline{\phantom{\alpha}}\!\!\!\gamma}
\def\um{\underline{\mu}} \def\un{\underline{\nu}}
\def\ud{\underline\delta}
\def\ue{\underline\epsilon}
\def\una{\underline a}\def\unA{\underline A}
\def\unb{\underline b}\def\unB{\underline B}
\def\unc{\underline c}\def\unC{\underline C}
\def\und{\underline d}\def\unD{\underline D}
\def\une{\underline e}\def\unE{\underline E}
\def\unf{\underline{\phantom{e}}\!\!\!\! f}\def\unF{\underline F}
\def\unm{\underline m}\def\unM{\underline M}
\def\unn{\underline n}\def\unN{\underline N}
\def\unp{\underline{\phantom{a}}\!\!\! p}\def\unP{\underline P}
\def\unq{\underline{\phantom{a}}\!\!\! q}
\def\unQ{\underline{\phantom{A}}\!\!\!\! Q}
\def\unH{\underline{H}}
\def\ul{\underline}

\def\As {{A \hspace{-6.4pt} \slash}\;}
\def\bs {{b \hspace{-6.4pt} \slash}\;}
\def\Ds {{D \hspace{-6.4pt} \slash}\;}
\def\ds {{\del \hspace{-6.4pt} \slash}\;}
\def\ss {{\s \hspace{-6.4pt} \slash}\;}
\def\ks {{ k \hspace{-6.4pt} \slash}\;}
\def\ps {{p \hspace{-6.4pt} \slash}\;}
\def\pas {{{p_1} \hspace{-6.4pt} \slash}\;}
\def\pbs {{{p_2} \hspace{-6.4pt} \slash}\;}

\def\Fh{\hat{F}}
\def\Vh{\hat{V}}
\def\Xh{\hat{X}}
\def\ah{\hat{a}}
\def\xh{\hat{x}}
\def\yh{\hat{y}}
\def\ph{\hat{p}}
\def\xih{\hat{\xi}}

\def\psit{\tilde{\psi}}
\def\Psit{\tilde{\Psi}}
\def\tht{\tilde{\th}}

\def\At{\tilde{A}}
\def\Qt{\tilde{Q}}
\def\Rt{\tilde{R}}
\def\Nt{\tilde{N}}

\def\CT{\tilde{C}}
\def\BT{\tilde{B}}
\def\ta{\tilde{a}}
\def\st{\tilde{s}}
\def\ft{\tilde{f}}
\def\pt{\tilde{p}}
\def\qt{\tilde{q}}
\def\vt{\tilde{v}}
\def\nt{\tilde{n}}
\def\tu{\tilde{u}}

\def\delb{\bar{\partial}}
\def\bz{\bar{z}}
\def\bD{\bar{D}}
\def\bB{\bar{B}}

\def\bk{{\bf k}}
\def\bl{{\bf l}}
\def\bp{{\bf p}}
\def\bq{{\bf q}}
\def\br{{\bf r}}
\def\bx{{\bf x}}
\def\by{{\bf y}}
\def\bR{{\bf R}}
\def\bV{{\bf V}}

\def\d{\delta}\def\D{\Delta}\def\ddt{\dot\delta}

\def\p{\partial} \def\del{\partial}
\def\xx{\times}
\def\uno{\mbox{1 \kern-.59em {\rm l}}}

\def\trp{^{\top}}
\def\inv{^{-1}}
\def\dag{{^{\dagger}}}

\def\pr{\prime}
\def\rar{\rightarrow}
\def\lar{\leftarrow}
\def\lrar{\leftrightarrow}
\def\rt{\right}

\begin{document}
\title{Classical Aspects of Higher Spin Topologically Massive Gravity}
\author{Bin Chen$^{1,2}$\footnote{Email:bchen01@pku.edu.cn}\hs{2ex}Jiang Long$^1$\footnote{Email:lj301@pku.edu.cn}\hs{2ex}and
Jian-dong Zhang$^1$\footnote{Email:mksws@pku.edu.cn}\\
\small{$^1$Department of Physics,
and State Key Laboratory of
Nuclear Physics and Technology,}\\
\small{Peking University, Beijing 100871, P.R. China}\\
\small{$^2$Center for High Energy Physics,
Peking University, Beijing 100871, P.R. China}\\
}
\date{\today}
\maketitle

\begin{abstract}
We study the classical solutions of three dimensional topologically massive gravity (TMG) and its higher spin generalization, in the first order formulation. The action of higher spin
TMG has been proposed in \cite{Chen:2011yx} to be of a Chern-Simons-like form. The equations of motion are more complicated than the ones in pure higher spin AdS$_3$ gravity, but are still tractable. As all the solutions in higher spin gravity are automatically the solutions of higher spin TMG, we focus on other solutions. We manage to find the AdS pp-wave solutions with higher spin hair, and find that the nonvanishing higher spin fields may or may not modify the pp-wave geometry. In order to discuss the warped spacetime, we introduce the notion of special Killing vector, which is defined to be the symmetry on the frame-like fields. We reproduce various warped spacetimes of TMG in our framework, with the help of special Killing vectors.
 \end{abstract}

\newpage

\section{Introduction}

 Three dimensional topologically massive gravity(TMG)\cite{Deser:1982vy,Deser:1981wh} provides an interesting modification of Einstein's general relativity. Besides the Einstein-Hilbert term and  the cosmological constant $\L$, the action of TMG theory includes also a gravitational Chern-Simons term
\bea \label{z5}
&&I_{TMG}=I_{EH}+I_{CS}  \nn\\
&&\phantom{I_{TMG}}
=\frac{1}{16\pi G}\int d^3x \sqrt{-g} \lt[ R-2\L  +\frac{1}{2\m}\e^{\l\m\n}\G^\r_{\l\s}
                             \lt( \p_\m\G^\s_{\r\n} +\frac{2}{3}\G^\s_{\m\t}\G^\t_{\n\r} \rt) \rt].\nn
\eea
Unlike the usual three-dimensional pure gravity which has no local degree of freedom, there is generically a massive propagating degree of freedom in TMG. The equation of motion in TMG is a third order differential equation, involving the Cotton tensor. As a result, the usual solutions of the  Einstein gravity with/without a cosmological constant, which must be of a Einstein metric, are also the solutions of the TMG theory. Moreover there are new solutions which has nonvanishing Cotton tensor\cite{solution}. For the TMG with a cosmological constant, these solutions include the spacelike, timelike, null warped spacetimes, AdS pp-wave and also regular black hole solutions. Though the full classification of the solutions of the TMG theory has not been achieved, much progress has been made in the past few years. Especially, in \cite{Chow:2009km,Chow:2009vt,ertl}, the Petrov-Segre classification of 3D TMG has been discussed carefully.

On the other hand, there has been much progress on the higher spin AdS$_3$ gravity in the past two years. In 3D, the higher spin gravity is simpler and could be put into a Chern-Simons form even for finite spin. In a remarkable paper\cite{Theisen:2010}, the authors showed that the high spin AdS$_3$ gravity theory with spin up to $n$ could be written as a Chern-Simons gravity theory with
$SL(n,R)\times SL(n,R)$ gauge group.
 By imposing appropriate boundary condition and gauge choice, they showed that the asymptotic symmetry group of higher spin AdS$_3$ gravity is a classical $W_n$ symmetry with the same central charge in the pure AdS$_3$ gravity found in \cite{Brown:1986nw}. This indicates that the higher spin AdS$_3$ gravity is holographically dual to a conformal field theory with $W_n$ symmetry. Similar observation has been made in \cite{Henneaux:2010xg} starting from the $n\to \infty$ theory. 

Like spin-2 graviton in pure 3D gravity, the higher spin fields in AdS$_3$ are not dynamical neither. It is quite interesting to consider the generalization of TMG to the higher spin case. It was actually in 1988 that the Chern-Simons-like coupling for spin-3 field in flat spactime has been proposed\cite{Damour:1987vm}, based on the gauge symmetry consideration. The similar Chern-Simons-like terms for higher spin fields in AdS$_3$ were suggested in \cite{Sahoo 1:2011}. The full action describing the higher spin topologically fields coupled to the AdS$_3$ gravity has been proposed in \cite{Chen:2011vp,Chen:2011yx}. As in usual higher spin AdS$_3$ gravity, it is constructed in terms of the frame-like fields. In such first order formulation, it is remarkable that the action of the higher spin topologically massive gravity could be written as a Chern-Simons-like action as well. Besides two Chern-Simons action with different levels, there is another extra term which leads to torsion-free conditions by a Lagrangian multiplier. The presence of the Lagrangian multiplier field make the gauge symmetry not manifest. Fortunately, it was shown in \cite{Chen:2011yx}, in the AdS$_3$ vacuum the gauge symmetry is reproduced and plays an essential role to relate the framelike fields to the physical fields. The analysis of linearized action of the fluctuations around  the AdS$_3$ vacuum
shows that the equations of motion of the higher spin fields are the same as the ones obtained from the Frondal fields with higher spin Chern-Simons terms\cite{Sahoo 2:2011}.

In this paper, we would like to study the classical solutions of the higher spin topological massive gravity, in the first-order formulation. In the pure higher spin AdS$_3$ gravity, the classical solutions asymptotic to AdS$_3$ has been studied carefully in \cite{Gutperle:2011kf,Ammon:2011nk,Castro:2011iw}. In this case, the solutions are the gauge flat connections, under appropriate boundary conditions.\footnote{For the discussion of Non-AdS solutions in pure higher spin AdS$_3$ gravity, see \cite{Gary:2012ms}.} In the case of higher spin TMG, the equations of motion are much more involved. 
It makes things worse that we do not know how to impose appropriate boundary conditions for the warped spacetimes in this formulation. Therefore in this paper, we take the strategy that we do not fix any boundary condition and solve the equations of motion directly. 
We manage to get a class of AdS pp-wave solutions with nontrivial higher spin fields. However, it turns out to be difficult to find the warped AdS$_3$ spacetime with nontrivial higher spin hair. Nevertheless, we find the classical solutions of 3D TMG theory in the first-formulation, corresponding to the spacelike, timelike, null warped AdS$_3$ spacetimes and also spacelike warped AdS$_3$ blackhole. Such investigation has not been carried out in the literature. This is another motivation behind our study.

 In our study, we develop the notion of special Killing vector(SKV), which is defined to be the symmetry generators acting on the frame-like fields via Lie-derivative. The special Killing vectors form a subalgebra of the isometry algebra. As SKV is also the symmetry acting on the gauge potentials, it help us to solve the equations of motion.
It turns out that the first-order differential equations of motion could be transformed to a set of matrix equations with appropriate SKV ansatz, which could be solved in principle.

Another remarkable fact from our study is that there are usually many solutions corresponding to one spacetime. For example, corresponding to the same spacelike warped AdS$_3$ spacetime, we find a two-parameter class of the solutions in the first-order formulation. And we also find the same spacelike warped AdS$_3$ spacetime appearing in different metric forms, due to the exchange symmetry in the matrix equations. Furthermore, we find the non-principal embedding of $SL(2,R)$ in $SL(3,R)$, which leads to different warped AdS$_3$ vacuum with vanishing higher spin fields.

The remaining parts of the paper are organized as follows. In Sect. 2, we review the HSTMG theory and its properties. In Sect. 3, we discuss the AdS pp-wave solutions with higher spin hair, after imposing appropriate boundary conditions to ensure asymptotically being AdS$_3$. In Sect. 4, we investigate the various warped AdS$_3$ spacetimes. We pay special attention to spacelike warped AdS$_3$ spacetime, though the techniques and results could be generalized to timelike, null warped cases as well. We end with some discussions in Sect. 5.

\section{Action and symmetry}

The action of higher spin topologically massive gravity in the first order formulation could be cast into a Chern-Simons-like form\cite{Chen:2011yx}
\be
S_{HSTMG}=\frac{k}{4\pi}[(1-\frac{1}{\mu l})S_{CS}[A]-(1+\frac{1}{\mu l})S_{CS}[\bar{A}]-\frac{1}{\mu}\int tr(\beta\wedge(F-\bar{F}))].
\ee
where
\be
S_{CS}[A]=\int tr(A\wedge dA+\frac{2}{3}A\wedge A\wedge A)
\ee
is proportional to the Chern-Simons action and
\be
F=dA+A\wedge A; \hs{3ex}\bar{F}=d\bar{A}+\bar{A}\wedge\bar{A}.
\ee
Note that our definition of the Chern-Simons action $S_{CS}[A]$ is slightly different from the conventional one since we have taken out the Chern-Simons level $k$. There are two other parameters in our action. The parameter $l$ is related to a negative cosmological constant $\Lambda=-1/l^2$ and is the radius of AdS vacuum. $\mu$ has the dimension of mass and it is proportional to the mass of the massive higher spin modes\cite{Chen:2011yx}. There is a one-form $\beta$ in the last term, which imposes the torsion free condition. Without it, the action becomes the sum of two pure Chern-Simons action with different levels. The gauge field $A$ and $\bar{A}$ are related to the vielbein $e$ and spin connection $\omega$ by $A=\omega+e/l, \bar{A}=\omega-e/l$ and they take value in some suitable Lie algebra.
If the gauge fields take value in $SL(2,R)$, the above action is equivalent to the one for the topologically massive gravity. Therefore, for the general gauge group $SL(n,R)$, the above action is a natural choice for higher spin topologically massive gravity. Actually, the study of the perturbations around AdS$_3$ vacuum in \cite{Chen:2011yx} shows that at the linearized level the action is equivalent to the one with higher spin Chern-Simons terms  suggested from gauge symmetry.

 Variating the action with respect to $A$,$\bar{A}$ and $\beta$ correspondingly, we find the following equations of motion\footnote{From now on, we set $l=1$.}
\bea
(1-\frac{1}{\mu })F&=&\frac{1}{2\mu}K;\label{eom1}\\
(1+\frac{1}{\mu })\bar{F}&=&\frac{1}{2\mu}\bar{K};\label{eom2}\\
F&=&\bar{F}. \label{eom3}
\eea
where we have defined the two-form $K$ and $\bar{K}$ as
\bea
K&\equiv&D_A\b=d\beta+\beta\wedge A+A\wedge\beta;\nn\\
\bar{K}&\equiv&D_{\bar A}\b=d\beta+\beta\wedge \bar{A}+\bar{A}\wedge\beta.
\eea
The last equation (\ref{eom3}) gives torsion-free condition for the fields of various spins and the first two equations (\ref{eom1}) and (\ref{eom2}) modifies the flatness condition of the gauge connections. However, to have a well-defined variation, we need to deal with the boundary terms appropriately. The variation of our action, including the boundary terms, is
\be
\delta S=\int_{M} (e.o.m)+\frac{k}{4\pi}\int_{\partial M}tr(\frac{1}{\mu}\beta-(1-\frac{1}{\mu})A)\wedge\delta A-\frac{k}{4\pi}\int_{\partial M}tr(\frac{1}{\mu}\beta-(1+\frac{1}{\mu})\bar{A})\wedge\delta \bar{A}.
\ee
There are two kinds of methods to cancel the boundary terms. The first one is to add some counter-terms to cancel them, 
while the second one is to put suitable boundary conditions on the fields. 
 In this
paper we always assume that we can have a well defined variational principal to get the equations of motion. In other words,
we regard the equations of motion as our starting point. Of course,
 if we are interested in the solutions that are asymptotic to AdS$_3$, we require the following reasonable condition that once we are near the boundary, the extra lagrangian multiplier should be asymptotic to zero
\be
\beta\to 0, \hs{3ex}\mbox{for}\ \rho\to\infty.
\ee
But one should note that these cannot be the whole boundary conditions.

As we have shown in \cite{Chen:2011yx}, for general $\beta$, the higher spin topologically massive gravity is invariant only under the infinitesimal gauge transformation
\be
\delta_{\Lambda}A=d\Lambda+[A,\Lambda],\ \delta_{\Lambda}\bar{A}=d\Lambda+[\bar{A},\Lambda],\ \delta_{\Lambda}\beta=[\beta, \Lambda].
\ee
which can be interpreted as generalized local Lorentz transformation. In our case, we ignore the possible subtlety of large gauge transformation. Then the finite gauge transformation of $\beta$ can be written as
\be
\beta\to U\beta U^{-1}\label{gt}
\ee
  We may use $\beta$ to classify the classical solutions of the theory.  The solutions of higher spin gravity with $\beta=0$ and $\beta\not=0$ are not equivalent. The reason is that once $\beta\not=0$, one cannot make a gauge transformation (\ref{gt}) to set it to zero. At the same time, one cannot make a coordinate transformation to make it to zero since $\beta$ is a coordinate invariant one-form. This distinction is important once one find a ``new" solution, one can check the one-form $\beta$ to see whether it is really a new solution.

Note that the point $\beta=0$ is quite special since it is the point where the gauge symmetry is enlarged. In this case, the equations of motion show that the gauge potentials should be pure gauge, which is exactly the case in higher spin AdS$_3$ gravity. In other words, all the solutions in higher spin AdS$_3$ gravity are automatically the solutions of higher spin TMG. 

In this paper, we would like to discuss the nontrivial solution with $\b\neq 0$. As we are only interested in searching for the solutions, we do not impose any boundary conditions in advance. For principal embedding, once we solve the equations of motion (\ref{eom1}-\ref{eom3}), we may read the frame-like fields and obtain the metric and higher spin fields via the relations
\be
g_{\mu\nu}=\frac{1}{\epsilon_N^{(2)}}tr (e_{\mu}\cdot e_{\nu})\label{metric}
\ee
\be
\phi_{\mu_1\mu_2\mu_3}=\frac{1}{\epsilon_N^{(3)}}tr (e_{(\mu_1}\cdots e_{\mu_3)})\label{hs}
\ee
and similar expressions for other higher spin fields. Our definition of the metric and higher spin fields in terms
of veilbein is the same as pure higher spin gravity. The reason are:
\begin{enumerate}
\item  The physical fields ($\phi_{\mu_1\cdots\mu_s}$) must be local lorentz invariant.
\item  Once $\beta\to 0$, we need to go back to the corresponding definition in higher spin gravity.
\end{enumerate}
It is easy to show that the above definition satisfy such
conditions. It is interesting to see whether such kind of constraints forces us to give the unique definition.

\section{Higher Spin AdS pp-wave Solutions}

In this section we want to discuss higher spin AdS pp-wave solutions. These solutions are much like the higher spin realization of AdS pp-wave solutions. Since the equations of motion for $\mu>1$ and $\mu=1$ are quite different, we consider the $\mu>1$ case firstly. 
As we do not want to deform the conventional higher spin gravity much, we require
\be
\beta\to 0, \hs{3ex}{\mbox for}\ \rho\to\infty\label{bc}
\ee
and we also assume that
\be
A_{\rho}=L_0,\ \bar{A}_{\rho}=-L_0,\ \beta_{\rho}=0.\label{ansatz}
\ee
Note that such requirements should be regarded as part of our ansatz in contrast to the higher spin gravity case, in which one can always choose a gauge to set the solution to obey (\ref{ansatz}). In our case, we cannot do this gauge fixing procedure so (\ref{ansatz}) is just part of ansatz which will turn out to be consistent. Under this ansatz, one can solve the $\rho +$ and $\rho -$ components of the equations of motion. The results are
\bea
\beta_{+}&=&\exp(-\mu\rho L_0)X_+(z,\bar{z})\exp(\mu\rho L_0);\nonumber\\
A_+&=&\exp(-\rho L_0)a_+(z,\bar{z})\exp(\rho L_0)+\frac{1}{2(\mu-1)}\beta_+;\nonumber\\
\bar{A}_{+}&=&\exp(\rho L_0)\bar{a}_+(z,\bar{z})\exp(-\rho L_0)+\frac{1}{2(\mu+1)}\beta_+
\eea
and
\bea
\beta_{-}&=&\exp(-\mu\rho L_0)X_-(z,\bar{z})\exp(\mu\rho L_0);\nonumber\\
A_-&=&\exp(-\rho L_0)a_-(z,\bar{z})\exp(\rho L_0)+\frac{1}{2(\mu-1)}\beta_-;\nonumber\\
\bar{A}_{-}&=&\exp(\rho L_0)\bar{a}_-(z,\bar{z})\exp(-\rho L_0)+\frac{1}{2(\mu+1)}\beta_-
\eea
Note that the above solutions must satisfy the constraints from the $+-$ components of the equation of motion
\be
2(\mu-1)F_{+-}=K_{+-};\ 2(\mu+1)\bar{F}_{+-}=\bar{K}_{+-};\ F_{+-}=\bar{F}_{+-}.\label{cons}
\ee
To solve the above equations of motion is still quite difficult, so we further make the following ansatz
\be
A_-=0;\ \beta_-=0\label{inter}
\ee
to simplify the equations of motion. If we further require that all the fields are not independent of coordinate $\bar{z}$, and take the condition (\ref{bc}) into account, then for the gauge group $SL(2,R)$ we find the solution
\bea
\beta_+&=&\beta(z)L_{-1}\exp(-\mu\rho);\nonumber\\
A_{+}&=&\exp(\rho)L_1+\mathcal{L}(z)\exp(-\rho)L_{-1}+\frac{1}{2(\mu-1)}\beta(z)\exp(-\mu\rho)L_{-1};\nonumber\\
\bar{A}_+&=&\frac{1}{2(\mu+1)}\beta(z)\exp(-\mu\rho)L_{-1}.\nonumber\\
\beta_-&=&0;\label{sol1}\\
A_-&=&0;\nonumber\\
\bar{A}_-&=&\exp(\rho)L_{-1}.\nonumber
\eea
The solution (\ref{sol1}) plus (\ref{ansatz}) give us the metric via (\ref{metric})
\bea
ds^2=d\rho^2+[-\mathcal{L}(x^+)-\frac{1}{\mu^2-1}\beta(x^+)\exp(1-\mu)\rho](dx^+)^2+e^{2\rho} dx^+dx^-.\label{adsppwave}
\eea
This is exactly the AdS pp-wave solution found in \cite{Gibbons:2008vi}\footnote{For other forms of AdS-pp wave and other references on AdS pp-wave solution, see the appendix A.3 in \cite{Chow:2009km}.}. Note that we have replaced $z(\bar{z})$ with $x^+(x^-)$ to match the usual conventions of pp-wave solution. We can redefine $\beta(x^+)$ to absorb the factor $\frac{1}{\mu^2-1}$ which appears in the metric. We can do the same thing in the following.

For the gauge group $SL(n,R)$, one kind of higher spin AdS pp-wave solutions is of the following form
\bea
\beta_{+}&=&\sum_{s=2}^{n}\beta_{(s-1)}(z)W^{s}_{-(s-1)}\exp(-(s-1)\mu\rho);\nonumber\\
A_{+}&=&\exp(\rho)L_1+\sum_{s=2}^{n}\mathcal{W}_{(s-1)}(z)\exp(-(s-1)\rho)W^{s}_{-(s-1)}+\frac{1}{2(\mu-1)}\beta_+;\nonumber\\
\bar{A}_+&=&\frac{1}{2(\mu+1)}\beta_+;\hs{3ex}\bar{A}_-=\exp(\rho)L_{-1};\hs{3ex}\bar{A}_{\rho}=-L_0;\nonumber\\
\beta_-&=&0;\hs{3ex}\beta_{\rho}=0;\nonumber\\
A_-&=&0;\hs{3ex}A_{\rho}=L_0.\label{sol2}
\eea
Note that we have used the convention of principal embedding and the identification $W^2_{-1}=L_{-1}, W^2_1=L_1$. It is straightforward to  translate these solutions to the metric-like fields. For the special case $n=3$, we have
\be
ds^2=d\rho^2+[-\mathcal{W}_1(x^+)-\frac{1}{\mu^2-1}\mathcal{\beta}_1(x^+)e^{(1-\mu)\rho}](dx^+)^2+e^{2\rho}dx^+dx^-;
\ee
for the metric field and the nonzero components of spin-3 field are
\bea
\phi_{+++}=(\mathcal{W}_2(x^+)+\frac{\beta_2(x^+)}{\mu^2-1}e^{2(1-\mu)\rho}),
\eea
up to a normalization constant corresponding to spin-3 field.

 It is surprising  that the metric takes the same form as AdS pp-wave solution (\ref{adsppwave}) while the higher spin fields are non-vanishing at the same time! Namely, the turning-on of the higher spin hairs does not affect the the AdS pp-wave spacetime geometry. 
However, for more general solutions, this situation changes. We find that when $\beta_{+},\beta_-,\beta_{\rho},A_-,A_{\rho},\bar{A}_+, \bar{A}_-,\bar{A}_{\rho}$ take the same form as (\ref{sol2}), $A_+$ have a large number of degrees of freedom. Actually, we only require that $a_+(z,\bar{z})$ is a $SL(n,R)$ matrix function of $z$. The more general solutions are
\bea
\beta_{+}&=&\sum_{s=2}^{n}\beta_{(s-1)}(z)W^{s}_{-(s-1)}\exp(-(s-1)\mu\rho);\nonumber\\
A_{+}&=&\exp(-\rho L_0)a(z)\exp(\rho L_0)+\frac{1}{2(\mu-1)}\beta_+;\nonumber\\
\bar{A}_+&=&\frac{1}{2(\mu+1)}\beta_+;\hs{3ex}\bar{A}_-=\exp(\rho)L_{-1};\hs{3ex}\bar{A}_{\rho}=-L_0;\nonumber\\
\beta_-&=&0;\hs{3ex}\beta_{\rho}=0;\nonumber\\
A_-&=&0;\hs{3ex}A_{\rho}=L_0.\label{sol3}
\eea
To see what happens we choose $n=3$, and set $a(z)=L_1+\mathcal{\k}_2(z)W^3_2$. Then the metric becomes
\bea
ds^2=d\rho^2+[-\mathcal{W}_1(x^+)-\frac{1}{\mu^2-1}\mathcal{\beta}_1(x^+)e^{(1-\mu)\rho}+\nonumber\\\mathcal{\k}_2(x^+) (\mathcal{W}_2(x^+)+\frac{1}{\mu^2-1}\mathcal{\beta}_2(x^+)e^{2(1-\mu)\rho})](dx^+)^2+e^{2\rho}dx^+dx^-;
\eea
Therefore we find that in this case the higher spin fields contribute to the AdS pp-wave geometry. For the spin-3 fields, the non-vanishing components include $\phi_{+++},\phi_{++-},\phi_{+--}$.
The analysis of the $\mu=1$ case  goes parallel to the $\mu>1$ case. We give the final result as
\bea
\beta_+&=&\sum_{s=2}^{N}\beta_{(s-1)}(z)W^{s}_{-(s-1)}\exp(-(s-1)\rho);\nonumber\\
A_+&=&\exp(\rho)L_1+\sum_{s=2}^{N}\mathcal{W}_{(s-1)}(z)\exp(-(s-1)\rho)W^{s}_{-(s-1)}+\frac{1}{2}\rho\beta_+';\nonumber\\
\bar{A}_+&=& \frac{1}{4}\beta_+;\hs{3ex}\bar{A}_-=\exp(\rho)L_{-1};\hs{3ex}\bar{A}_{\rho}=-L_0.\nonumber\\
\beta_-&=&0;\hs{3ex}\beta_{\rho}=0;\nonumber\\
A_-&=&0;\hs{3ex}A_{\rho}=L_0;\label{sol4}
\eea
where $\beta_+'$ denotes $\partial_{\rho}\beta_+$.  We can generalize this solution by using the same method from (\ref{sol2}) to (\ref{sol3}).\\

In the previous discussion, we have assumed (\ref{inter}) to simplify the constraints (\ref{cons}). However, there are other ways to simplify those constraints. Here we just give one such way. We can assume that the matrix functions $X_+,X_-, a_+,\bar{a}_+, a_-,\bar{a}_-$ in (16,17) to be constants, namely, independent of any coordinates. Then the constraints (\ref{cons}) become
\bea
[A_+, A_-]&=&[\bar{A}_+,\bar{A}_-], \nn\nonumber\\ 2(\mu-1)[A_+,A_-]&=&[\beta_+,A_-]+[A_+,\beta_-],\nn\nonumber\\
2(\mu+1)[\bar{A}_+,\bar{A}_-]&=&[\beta_+,\bar{A}_-]+[\bar{A}_+,\beta_-].
\eea
These equations are independent of the choice of gauge group. If we choose $SL(2,R)$ and assume that
\bea
X_{+}&=&x_1 L_1+x_{-1}L_{-1},\nn\nonumber\\
X_-&=&y_1 L_1+y_{-1}L_{-1},\nn\nonumber\\
a_+&=&a_1 L_1+a_{-1}L_{-1},\nn\nonumber\\
\bar{a}_+&=&\bar{a}_1 L_1+\bar{a}_{-1}L_{-1},\nn\nonumber\\
a_-&=&b_1 L_1+b_{-1}L_{-1},\nn\nonumber\\
\bar{a}_-&=&\bar{b}_1 L_1+\bar{b}_{-1}L_{-1},
\eea
we find that there are many solutions which leads to well defined metric. For example, one kind of solution is
\bea
\{a_1, a_{-1}, \bar{a}_1, \bar{a}_{-1}, b_1, b_{-1}, \bar{b}_1, \bar{b}_{-1}, x_1, x_{-1}, y_1, y_{-1}\}=\nn\nonumber\\
\{\frac{b_1 x_{-1}}{y_{-1}}, 0, 0, \bar{a}_{-1}, b_1, 0, 0, \bar{b}_{-1}, 0, x_{-1}, 0, y_{-1}\},
\eea
in which there are five undetermined parameters. The corresponding nonvanishing metric components are
\bea
g_{++}&=&-\frac{b_1 x_{-1}}{y_{-1}}(-\bar{a}_{-1}e^{2\rho}+x_{-1}e^{(1-\mu)\rho});\nn\nonumber\\
g_{+-}&=&g_{-+}=-\frac{1}{2}b_1 (-\bar{a}_{-1}e^{2\rho}+x_{-1}e^{(1-\mu)\rho})-\frac{b_1 x_{-1}}{2y_{-1}}(-\bar{b}_{-1}e^{2\rho}+y_{-1}e^{(1-\mu)\rho});\nn\nonumber\\
g_{--}&=&b_1(\bar{b}_{-1}e^{2\rho}-y_{-1}e^{(1-\mu)\rho});\nn\nonumber\\
g_{\rho\rho}&=&1
\eea
Note that we have replaced $\frac{x_{-1}}{\mu^2-1}(\frac{y_{-1}}{\mu^2-1})$ to $x_{-1}(y_{-1})$. Certainly, we may choose the gauge group to be arbitrary $SL(n,R)$.

\section{Warped $AdS_3$ spacetime}

In 3D topologically massive gravity, there are other solutions besides the AdS pp-wave solutions discussed in the last section. Among them, the warped AdS$_3$ spacetimes are of particular interests. These spacetimes could be considered as the U(1)-fibred AdS$_2$ with a warping factor, so they have isometry group $SL(2,R) \times U(1)$. More interestingly, it has been conjectured that for spacelike warped AdS$_3$, there exists holographically 2D CFT description under appropriate boundary conditions\cite{Andy08}. Therefore it would be very interesting to discuss such spacetimes in our framework.

As the warped spacetimes have much isometry, it would be nice to see if the Killing symmetry could be helpful to find the solution. In the next subsection, we will introduce the notion  of special Killing symmetry. Then in the following subsections we use it to find the solutions corresponding to the warped spacetimes.

\subsection{Special Killing Vector}

In the usual formulation of gravity, the Killing symmetry is important to
make metric ansatz  to solve the Einstein's equations of motion. A suitable assumption of the symmetry of the solution always simplify the equations of motion.  In the first order formulation of the gravity, it is frame-like fields which appear in the Lagrangian and the equations of motion. It would be interesting to generalize the notion of Killing symmetry to the frame-like fields.  The key observation is that once
\be
\delta_{\hat{\xi}}e_{\mu}\equiv\mathcal{L}_{\hat{\xi}}e_{\mu}=\xi^\l\p_\l e_\m+\p_\m \xi^\l e_\l=0,\label{SKV}
\ee
then the metric must satisfy the Killing equation
\be
\delta_{\hat{\xi}}g_{\mu\nu}=\delta_{\hat{\xi}}tr(e_{\mu}\cdot e_{\nu})=0
\ee
then the vector $\hat{\xi}$ must be a Killing vector. A vector which satisfies (\ref{SKV}) is called Special Killing Vector(SKV) by us. Note that a Killing vector is not necessarily a SKV. Some properties of SKV are stated in the following\footnote{In this paper we focus on three dimension, the generalization to other dimensions should be straightforward.}:
\begin{enumerate}
\item If $\hat{\xi}$ and $\hat{\eta}$ are two SKVs and k is a constant number, then $\hat{\xi}+\hat{\eta}$, $k\hat{\xi}$ and $[\hat{\xi}, \hat{\eta}]$ are  SKVs, too.
\item The set of SKV forms an algebra which is a subalgebra of the whole Killing symmetry algebra.
\item For arbitrary continuous coordinate transformation $x^{\mu}\to x^{'\mu}$, $e_{\mu}\to e'_{\mu}$, $\xi^{\mu}\to\xi^{'\mu}$,  if $\delta_{\xi}e_{\mu}=0$, then $\delta_{\xi'}e'_{\mu}=0$. In other words, SKV is coordinate independent;
\item Under a global Lorentz transformation generated by $\Lambda$, $\delta e=[e,\Lambda]$, $\delta_{\xi}e_{\mu}=0$. However, under an arbitrary local Lorentz transformation, $\delta_{\xi}e_{\mu}$ can become nonzero.
    \end{enumerate}

Actually, due to the fact that Lie-derivative and differential operator are commutative, from the Cartan's equation the torsion-free spin-connection 1-form satisfies
\be
\mathcal{L}_{\hat{\xi}}{\hat \o}=0,
\ee
provided that the $\hat{\xi}$ is SKV. Therefore, we have
\be
\mathcal{L}_{\hat{\xi}}A=0, \hs{3ex}\mathcal{L}_{\hat{\xi}}{\bar A}=0.\label{SKVA}
\ee
From the equations of motion, it is natural to set
\be
\mathcal{L}_{\hat{\xi}}\b=0.\label{SKVB}
\ee
The relations (\ref{SKVA},\ref{SKVB}) are quite restrictive.

We have shown that the SKV is not local Lorentz invariant.  One resolution is that we only require that $\delta_{\xi}e_{\mu}=0$ is valid only in some special frames. Another resolution is to change the definition of SKV to make it local Lorentz invariant. To require $\delta_{\xi}g_{\mu\nu}=0$, the constraint $\delta_{\xi}e_{\mu}=0$ could be too strong. Actually, $\delta_{\xi}e_{\mu}=[e_{\mu},G(\xi)]$ is sufficient, where $G(\xi)$is an arbitrary zero-form  depending on the choice of vector. One can verify easily that this modification does not change the first three properties of the SKV and moreover the concept of SKV becomes local Lorentz invariant. However, though this modification sounds attractive, it may break the argument which leads to (\ref{SKVA}) and (\ref{SKVB}). In the following discussion, we do not need to take into account of the local Lorentz invariance, so that we still use Eq. (\ref{SKV}) as the definition of the SKV.

\subsection{Timelike warped AdS$_3$}

For the timelike warped AdS$_3$, it has isometry $SL(2,R)_L \times U(1)$. The generators
of $SL(2,R)_L$ can be parameterized as
\bea
{\cal J}_1&=& \frac{\sinh\phi}{\cosh \r}\p_\t-\cosh\phi \p_\r +2\tanh\r\sinh\phi\p_\phi \nn\\
{\cal J}_2&=&\p_\phi\nn\\
{\cal J}_0&=&-\frac{\cosh\phi}{\cosh\r}\p_\t+\sinh\phi \p_\r-\tanh\r\cosh\phi\p_\phi,\nn
\eea
which satisfy the commutation relations
\be
[{\cal J}_1,{\cal J}_2]={\cal J}_0,\hs{3ex}[{\cal J}_0,{\cal J}_1]=-{\cal J}_2,\hs{3ex}[{\cal J}_0,{\cal J}_2]={\cal J}_1.
\ee
If we requires that the above $SL(2,R)_L$ algebra generators are the SKV's, then the gauge potentials and field $\b$ should satisfy (\ref{SKVA},\ref{SKVB}) with $\hat \xi$ being ${\cal J}_i$'s, we find that
\bea
A&=&i C_0d\t+(-i\sin\t C_1+\cos \t C_2)d\r+i(\sinh\r C_0+\cosh\r(\cos\t C_1-i\sin\t C_2))d\phi, \nn\\
{\bar A}&=&i {\bar C}_0d\t+(-i\sin\t {\bar C}_1+\cos \t{\bar C}_2)d\r+i(\sinh\r {\bar C}_0+\cosh\r(\cos\t {\bar C}_1-i\sin\t{\bar C}_2))d\phi, \nn\\
\b &=&i B_0d\t+(-i\sin\t B_1+\cos \t B_2)d\r+i(\sinh\r B_0+\cosh\r(\cos\t B_1-i\sin\t B_2))d\phi, \nn
\eea
where $C_i,\bar{C}_i,B_i$ are constant matrices. Moreover, the equations of motion (\ref{eom1}-\ref{eom3}) lead to a set of equations
\bea
C_1-[C_0,C_2]&=&\bar{C}_1-[\bar{C}_0,\bar{C}_2],\nn\\
C_2-[C_0,C_1]&=&\bar{C}_2-[\bar{C}_0,\bar{C}_1],\nn\\
C_0-[C_1,C_2]&=&\bar{C}_0-[\bar{C}_1,\bar{C}_2],\nn\\
(1-\frac{1}{\m })(C_1-[C_0,C_2])&=&\frac{1}{2\m}(B_1-[B_0,C_2]+[B_2,C_0]),\nn\\
(1-\frac{1}{\m })(C_2-[C_0,C_1])&=&\frac{1}{2\m}(B_2-[B_0,C_1]+[B_1,C_0]),\nn\\
(1-\frac{1}{\m })(C_0-[C_1,C_2])&=&\frac{1}{2\m}(B_0-[B_1,C_2]+[B_2,C_1]),\nn\\
(1+\frac{1}{\m })(\bar{C}_1-[\bar{C}_0,\bar{C}_2])&=&\frac{1}{2\m}(B_1-[B_0,\bar{C}_2]+[B_2,\bar{C}_0]),\nn\\
(1+\frac{1}{\m })(\bar{C}_2-[\bar{C}_0,\bar{C}_1])&=&\frac{1}{2\m}(B_2-[B_0,\bar{C}_1]+[B_1,\bar{C}_0]),\nn\\
(1+\frac{1}{\m })(\bar{C}_0-[\bar{C}_1,\bar{C}_2])&=&\frac{1}{2\m}(B_0-[B_1,\bar{C}_2]+[B_2,\bar{C}_1]).\label{HA1}
\eea
 It is interesting that these set of equations are independent of the choice of group algebra. Hence it seems that the structure under these set of equations are quite rich. Note that we have 9 unknown matrix to be determined while at the same time we have the same number of equations, hence there is potential that we have nontrivial solutions beyond warped AdS$_3$ solutions, though the structure may prevent us giving an explicit result. One can also observe that the equations (\ref{HA1}) are invariant under an $SL(n,R)$ transformation
\be
C_{i}\to UC_{i}U^{-1}, \bar{C}_i\to U\bar{C}_i U^{-1}, B_{i}\to UB_{i}U^{-1}
\ee
This is related to the global Lorentz invariance of the SKVs.\\\\

For general gauge group, the above equations are hard to solve. Even for the pure gravity with gauge group $SL(2,R)$, one has to make ansatz to simplify the equations. It is tempting to make ansatz that $C_i,\bar{C}_i,B_i$ are proportional to $SL(2,R)$ generators $J_i$ respectively, namely
\bea
i C_0=a_0J_0,&i C_1=a_1 J_1,&C_2=a_2J_2,\nn\\
i \bar{C}_0=\bar{a}_0J_0,&i \bar{C}_1=\bar{a}_1 J_1,&\bar{C}_2=\bar{a}_2J_2,\nn\\
i B_0=u_0J_0,&i B_1=u_1 J_1,&B_2=u_2J_2.\nn
\eea
Here the $SL(2,R)$ generators satisfy
\be
[J_0, J_1]=-J_2, \hs{3ex}[J_1, J_2]=J_0,\hs{3ex} [J_2, J_0]=-J_1.
\ee
The definition of $J_i, i=0,1,2$ are given in the appendix. With this set of ansatz,
the gauge potential and $\b$ could be rewritten as
\bea
A=a_0\sigma_0 J_0+a_1 \sigma_1 J_1+a_2\sigma_2 J_2\nn\\
\bar{A}=\bar{a}_0\sigma_0 J_0+\bar{a}_1 \sigma_1 J_1+\bar{a}_2\sigma_2 J_2\nn\\
\beta=u_0\sigma_0 J_0+u_1 \sigma_1 J_1+u_2\sigma_2 J_2\label{Asigma}
\eea
where
\bea
\sigma_0&\equiv& d\tau+\sinh \rho d\phi\nn\\
\sigma_1&\equiv &-\sin \tau d\rho+\cos \tau \cosh \rho d\phi\nn\\
\sigma_2&\equiv &\cos \tau d\rho+ \sin \tau \cosh \rho d\phi
\eea
satisfy the identities
\be
d\sigma_0=-\sigma_1\wedge\sigma_2,\hs{3ex} d\sigma_1=\sigma_2\wedge\sigma_0, \hs{3ex} d\sigma_2=\sigma_0\wedge\sigma_1.
\ee
The field strengths are now
\bea
F&=&(-a_0+a_1a_2)\sigma_1\wedge\sigma_2J_0+(a_1-a_0a_2)\sigma_2\wedge\sigma_0J_1+(a_2-a_0a_1)\sigma_0\wedge\sigma_1J_2,\nn\\
\bar{F}&=&(-\bar{a}_0+\bar{a}_1\bar{a}_2)\sigma_1\wedge\sigma_2J_0+(\bar{a}_1-\bar{a}_0\bar{a}_2)\sigma_2\wedge\sigma_0J_1+(\bar{a}_2-\bar{a}_0\bar{a}_1)\sigma_0\wedge\sigma_1J_2,\nn\\
K&=&(-u_0+u_1a_2+u_2a_1)\sigma_1\wedge\sigma_2J_0\nn\\
& & +(u_1-u_0a_2-u_2a_0)\sigma_2\wedge\sigma_0J_1+(u_2-u_0a_1-u_1a_0)\sigma_0\wedge\sigma_1J_2,\nn\\
\bar{K}&=&(-u_0+u_1\bar{a}_2+u_2\bar{a}_1)\sigma_1\wedge\sigma_2J_0\nn\\
& & +(u_1-u_0\bar{a}_2-u_2\bar{a}_0)\sigma_2\wedge\sigma_0J_1+(u_2-u_0\bar{a}_1-u_1\bar{a}_0)\sigma_0\wedge\sigma_1J_2. \label{eomsigma}
\eea
We further make ansatz that $a_1=a_2,\bar{a}_1=\bar{a}_2, u_1=u_2$, then we can solve the equations and obtain
\bea
a_0=-\frac{(\mu-3)(\mu+9)}{\mu^2+27},& &  a_1=\frac{|\mu-3|}{\sqrt{\mu^2+27}},\nn\\
\bar{a}_0=-\frac{(\mu-9)(\mu+3)}{\mu^2+27}, & & \bar{a}_1=\frac{\mu^2-9}{|\mu-3|\sqrt{\mu^2+27}},\nn\\
u_0=-\frac{8(\mu-3)\mu(\mu+3)}{3(\mu^2+27)},& & u_1=\frac{2|\mu-3|(\mu+3)}{3\sqrt{\mu^2+27}}.
\eea
From the definition of the metric,
\be
ds^2=2tr(e\otimes e)=\frac{1}{2}(a_0-\bar{a}_0)^2\sigma_0^2tr(J_0^2)
+\frac{1}{2}(a_1-\bar{a}_1)^2\sigma_1^2tr(J_1^2)+\frac{1}{2}(a_2-\bar{a}_2)^2\sigma_2^2tr(J_2^2),\nn
\ee
we find that
\be
ds^2=\frac{1}{\nu^2+3}[d\rho^2+\cosh \rho^2 d\phi^2-\frac{4\nu^2}{\nu^2+3}(d\tau+\sinh \rho d\phi)^2]\label{timeads}
\ee
where we have introduced a parameter $\nu=\frac{\mu}{3}$. This is exactly the timelike warped AdS$_3$ spacetime. It certainly has an isometry $SL(2,R)$ generated by SKV's, and also another Killing symmetry as the translation along $\t$. Note that the $U(1)$ Killing symmetry generated by $\p_\t$ is not a SKV. In a sense this symmetry is emergent from our solution.

From the equations of motion (\ref{eomsigma}), it is easy to find
the exchange symmetry among $a_0,a_1,a_2$. This suggests that we may make ansatz that $a_0=a_1$ etc. or $a_0=a_2$ etc.. Then we can find another solutions in the former case
\bea
ds^2&=&-\frac{1}{\nu^2+3}(d\tau+\sinh\rho d\phi)^2+\frac{1}{\nu^2+3}(-\sin\tau d\rho+\cos\tau\cosh\rho d\phi)^2\nn\\
& &+\frac{4\nu^2}{(\nu^2+3)^2}(\cos\tau d\rho+\sin\tau\cosh\rho d\phi)^2.
\eea
At first looking, the metric seems different from the one in (\ref{timeads}). But actually
both of them describe the same spacetime.

\subsection{Spacelike Warped $AdS_3$}

For the spacelike warped AdS$_3$, it has isometry group $SL(2,R)_R\times U(1)$. The $SL(2,R)_R$ group are generated by the
\be
\tilde{\cal J}_1=\sin\tau\tanh\rho\partial_{\tau}-\cos\tau\partial_{\rho}+\frac{\sin{\tau}}{\cosh{\rho}}\partial_{\phi};
\ee
\be
\tilde{\cal J}_2=-\cos\tau\tanh\rho\partial_{\tau}-\sin\tau\partial_{\rho}-\frac{\cos\tau}{\cosh\rho}\partial_{\phi};
\ee
\be
\tilde{\cal J}_0=\partial_{\tau}.
\ee
And  $U(1)$ is generated by
\be
\tilde{\cal J}=\partial_{\phi}
\ee

As in the timelike case, we expect that the above $SL(2,R)$ is the SKV of the spacelike warped AdS$_3$. Then from (\ref{SKVA},\ref{SKVB}), the possible form of gauge potentials and $\b$ are
\bea
A&=&\CT_0d\phi+(\sinh\r \CT_0+\cosh\r(\cosh\phi \CT_1+\sinh\phi \CT_2))d\t-(\sinh\phi \CT_1+\cosh\phi \CT_2)d\r \nn\\
\bar{A} &=&\bar{\CT}_0d\phi+(\sinh\r \bar{\CT}_0+\cosh\r(\cosh\phi \bar{\CT}_1+\sinh\phi \bar{\CT}_2))d\t-(\sinh\phi \bar{\CT}_1+\cosh\phi \bar{\CT}_2)d\r \nn\\
\b&=&\BT_0d\phi+(\sinh\r \BT_0+\cosh\r(\cosh\phi \BT_1+\sinh\phi \BT_2))d\t-(\sinh\phi \BT_1+\cosh\phi \BT_2)d\r \nn
\eea
where $\CT_i,\bar{\CT}_i, \BT_i$ are matrix-valued constants.   Moreover, the equations of motion lead to the following relations
\bea
\CT_1+[\CT_0,\CT_2]&=&\bar{\CT}_1+[\bar{\CT}_0,\bar{\CT}_2],\nn\\
\CT_2+[\CT_0,\CT_1]&=&\bar{\CT}_2+[\bar{\CT}_0,\bar{\CT}_1],\nn\\
\CT_0+[\CT_1,\CT_2]&=&\bar{\CT}_0+[\bar{\CT}_1,\bar{\CT}_2],\nn\\
(1-\frac{1}{\m })(\CT_1+[\CT_0,\CT_2])&=&\frac{1}{2\m}(\BT_1+[\BT_0,\CT_2]-[\BT_2,\CT_0])\nn\\
(1-\frac{1}{\m })(\CT_2+[\CT_0,\CT_1])&=&\frac{1}{2\m}(\BT_2+[\BT_0,\CT_1]-[\BT_1,\CT_0])\nn\\
(1-\frac{1}{\m })(\CT_0+[\CT_1,\CT_2])&=&\frac{1}{2\m}(\BT_0+[\BT_1,\CT_2]-[\BT_2,\CT_1])\nn\\
(1+\frac{1}{\m })(\bar{\CT}_1+[\bar{\CT}_0,\bar{\CT}_2])&=&\frac{1}{2\m}(\BT_1+[\BT_0,\bar{\CT}_2]-[\BT_2,\bar{\CT}_0])\nn\\
(1+\frac{1}{\m })(\bar{\CT}_2+[\bar{\CT}_0,\bar{\CT}_1])&=&\frac{1}{2\m}(\BT_2+[\BT_0,\bar{\CT}_1]-[\BT_1,\bar{\CT}_0])\nn\\
(1+\frac{1}{\m })(\bar{\CT}_0+[\bar{\CT}_1,\bar{\CT}_2])&=&\frac{1}{2\m}(\BT_0+[\BT_1,\bar{\CT}_2]-[\BT_2,\bar{\CT}_1]).\label{HA}
\eea

We may make ansatz that $\CT,\bar{\CT},\BT$'s are proportional to the generator $J_i$ of gauge group SL(2,R) algebra respectively, say
\be
\CT_0=i\ta_0J_0, \hs{3ex}\CT_1=\ta_1 J_1, \hs{3ex}\CT_2=\ta_2 J_2,
\ee
and similar to $\bar{\CT},\BT$'s. Then the gauge potentials and field $\b$ are of
the same forms as (\ref{Asigma}), with
 $\sigma_0,\sigma_1,\sigma_2$ being
\bea
\sigma_0&=&i(d\phi+\sinh \rho d\tau)\nn\\
\sigma_1&=&-i(\cosh\phi d\rho-\sinh \phi \cosh\rho d\tau)\nn\\
\sigma_2&=&-(\sinh \phi d\rho-\cosh \phi \cosh \rho d\tau).
\eea
We make similar ansatz as the timelike case: $\ta_1=\ta_2, \bar{\ta}_1=\bar{\ta}_2, \tu_1=\tu_2$, then we find that
\bea
\ta_0=\frac{(\mu-3)(\mu+9)}{\mu^2+27}, & & \ta_1=-\frac{\sqrt{-(\mu-3)^2}}{\sqrt{\mu^2+27}},\nn\\
\bar{\ta}_0=\frac{(\mu-9)(\mu+3)}{\mu^2+27}, & & \bar{\ta}_1=\frac{\mu^2-9}{\sqrt{-(\mu-3)^2}\sqrt{\mu^2+27}},\nn\\
\tu_0=\frac{8(\mu-3)\mu(\mu+3)}{3(\mu^2+27)}, & & \tu_1=-\frac{2\sqrt{-(\mu-3)^2}(\mu+3)}{3\sqrt{\mu^2+27}}.
\eea
Then the metric is
\be
ds^2=\frac{1}{\nu^2+3}[d\rho^2-\cosh^2 \rho d\tau^2+\frac{4\nu^2}{\nu^2+3}(d\phi+\sinh\rho d\tau)^2]\label{SWA}
\ee
where $\nu=\frac{\mu}{3}$. As in timelike case, there exists exchange symmetry in the equations of motion, which allows us to find other forms of gauge potentials leading to the same spacelike warped AdS$_3$. The details of the construction could be found in Appendix B.

More generally, all \(\CT_i, \bar{\CT}_i, \BT_i\) can be a combination of \(J_0, J_1, J_2\). First, we assume that \(\CT_i, \bar{\CT}_i, \BT_i\) have the similar form and proportional to each other: \(\bar{\CT}_i=k_i \CT_i\) and \(\BT_i=l_i \CT_i\). With this ansatz, we can fix the coefficients from the equations of motion
\bea
l_0=\frac{8\mu(\mu+3)}{3(\mu+9)},& l_1=\frac{2(\mu+3)}{3},& l_2=\frac{2(\mu+3)}{3},\nn\\
k_0=\frac{(\mu-9)(\mu+3)}{(\mu+9)(\mu-3))},&k_1=\frac{\mu+3}{\mu-3},&k_2=\frac{\mu+3}{\mu-3}\eea
or
\bea
l_1=\frac{8\mu(\mu+3)}{3(\mu+9)},& l_0=\frac{2(\mu+3)}{3},&l_2=\frac{2(\mu+3)}{3},\nn\\
k_1=\frac{(\mu-9)(\mu+3)}{(\mu+9)(\mu-3))},&k_0=\frac{\mu+3}{\mu-3},&k_2=\frac{\mu+3}{\mu-3}\eea
or
\bea
l_2=\frac{8\mu(\mu+3)}{3(\mu+9)},& l_1=\frac{2(\mu+3)}{3},&l_0=\frac{2(\mu+3)}{3},\nn\\
k_2=\frac{(\mu-9)(\mu+3)}{(\mu+9)(\mu-3))},&k_1=\frac{\mu+3}{\mu-3},&k_0=\frac{\mu+3}{\mu-3}.\eea
They are very similar to each other, with  the subscript exchanged. The solutions are consistent with the solutions got above.

Let us consider the first solution without losing the generality. After taking the $k_i$'s and $l_i$'s into the equations, we get
\begin{eqnarray}
\CT_1-\frac{{\mu}^2+27}{(\mu+9)(\mu-3)}[\CT_0,\CT_2]&=&0,\\
\CT_2-\frac{{\mu}^2+27}{(\mu+9)(\mu-3)}[\CT_0,\CT_1]&=&0,\\
\CT_0-\frac{\mu+9}{\mu-3}[\CT_1,\CT_2]&=&0.
\end{eqnarray}
We can set \(\CT_j=a^i_j J_i\), then we get algebraic equations for \(a^i_j\). There are only 9 unknown numbers, but the equations cannot be solved easily.
If we set two of \(a^i_j\) are zero, then according to the equations, the other two will be zero, too. So, there are only 5 unknown numbers, which could be determined from the equations. For example, if we set \(a^1_0=a^2_0=a^0_1=a^0_2=0\), the solutions can be written as the following with \(a^1_1\) being free
\[a^0_0=i\frac{(\mu-3)(\mu+9)}{\mu^2+27},~a^1_2=\sqrt{(a^{1}_1)^2+\frac{(\mu-3)^2}{\mu^2+27}},~a^2_1=-ia^1_2,~a^2_2=-ia^1_1. \]

If we set only one of \(a^i_j\) to be zero, for example, we set \(a^0_0=0\), the solutions have two free parameters \(a^1_0\) and \(a^0_1\)
\bea
a_2^0=\sqrt{(a_1^0)^2-\frac{1}{kq}},\hs{3ex}&&a_0^2=\sqrt{(a_0^2)^2-\frac{1}{k^2}}\nn\\
a_2^1=-k a_0^2 a_1^0,\hs{3ex}&&
a_2^2=k a_0^1 a_1^0,\nn\\
a_1^2=k a_0^1 a_2^0,\hs{3ex}&&
a_1^1=-k a_0^2 a_2^0
\eea
 where we have defined $k = \frac{\nu^2 + 3}{(\nu + 3) (\nu - 1)}, q = \frac{\nu + 3}{(\nu - 1)}$.
For the case of timelike, the solutions are similar with a change of sign.
Even though there are  two free parameters in the above solutions, they all lead to the same spacelike warped spacetime, as the two parameters get canceled  and do not appear in the explicit form of  the metric.

We have emphasized before that the matrix equations make sense for general gauge group. We may take the gauge group to be $SL(3,R)$, whose generators include the $SL(2,R)$ generators $L_{0,\pm 1}$ and the other generators $W_{\pm 2, \pm 1, 0}$. Let us first choose
\bea
\CT_0&=&a L_0,\nn\\
\CT_1&=&b_1 L_1+b_{-1}L_{-1},\nn\\
\CT_2&=&c_1 L_1+c_{-1}L_{-1}.
\eea
Then we find the one-parameter class of solutions characterized by $c_{-1}$  is
\bea
c_1 = -1/(4 c_{-1} k q),&&
b_{-1} = c_{-1}\nn\\
 b_1 =4 b_{-1} c_1^2 k q,& &
 a = -4 b_{-1} c_1 q,
 \eea
 where $k,q$ are defined as before. Then the metric are
 \be
 ds^2=\frac{1}{\nu^2+3}[d\rho^2-\cosh^2 \rho d\tau^2+\frac{4\nu^2}{\nu^2+3}(d\phi+\sinh\rho d\tau)^2]
 \ee
 and spin-3 fields are
 \be
 \phi_{\mu\nu\rho}\equiv0
 \ee
 There is no surprise since we just use the principal embedding of $SL(2,R)$ into $SL(3,R)$.
 On the other hand, as we know, we can choose another embedding like
 \bea
 \CT_0&=&a L_0;\nn\\
 \CT_1&=&b_2 W_2+b_{-2}W_{-2};\nn\\
 \CT_2&=&c_2 W_2+c_{-2}W_{-2};
 \eea
The one-parameter class of solutions characterized by $c_{-2}$ becomes
\bea
c_2 = 1/(16 c_{-2} k q),& & b_{-2} = c_{-2},\nn\\
b_2 = -16 b_{-2} c_2^2 k q,& & a = 8 b_{-2} c_2 q.
\eea
The corresponding metric becomes
\be
ds^2=\frac{1}{4(\nu^2+3)}[d\rho^2-\cosh^2 \rho d\tau^2+\frac{4\nu^2}{\nu^2+3}(d\phi+\sinh\rho d\tau)^2].
\ee
This kind of non-principal embedding changes the radius of warped $AdS_3$. So we see that in the warped spacetime solution, there exist non-principal
embedding as well.

\subsection{Warped black hole solution}

In 3D TMG theory, there exist warped black hole solutions.  They are locally warped spacetimes and  could be constructed by quotient identification of globally warped spacetimes. However, such quotient identification is useless in finding the solution in the Chern-Simons-like
theory of TMG. Nevertheless the singular coordinate transformations between warped black holes and warped spacetime are still useful for finding the solutions. It turns out the for the spacelike stretched black holes, whose metric is
\bea
ds^2&=&dt^2+\frac{dr^2}{(\n^2+3)(r-r_+)(r-r_-)}+(2\n r-\sqrt{r_+r_-(\n^2+3)})dtd\th \nn \\
  &&+\frac{r}{4}(3(\n^2-1)r+(\n^2+3)(r_++r_-)-4\n\sqrt{r_+r_-(\n^2+3)})d\th^2\nn
  \eea
  with $r\in [0,\infty],t\in [-\infty, \infty], \th\sim \th +2\pi$, the gauge potentials and $\b$ are respectively
  \bea
  A&=&\frac{\n^2+3}{2\n}(ia_0J_0)dt -\frac{a_1}{f(r)}(-J_1\sinh M+iJ_2\cosh M)dr \nn\\
  &&+\frac{\n^2+3}{2}\big(g(r)(ia_0J_0)+f(r)a_1(-J_1\cosh M+iJ_2\sinh M)\big)d\th, \nn\\
  \bar{A}&=&\frac{\n^2+3}{2\n}(i\bar{a}_0J_0)dt -\frac{\bar{a}_1}{f(r)}(-J_1\sinh M+iJ_2\cosh M)dr \nn\\
  &&+\frac{\n^2+3}{2}\big(g(r)(i\bar{a}_0J_0)+f(r)\bar{a}_1(-J_1\cosh M+iJ_2\sinh M)\big)d\th, \nn\\
  \b&=&\frac{\n^2+3}{2\n}(iu_0J_0)dt -\frac{u_1}{f(r)}(-J_1\sinh M+iJ_2\cosh M)dr \nn\\
  &&+\frac{\n^2+3}{2}\big(g(r)(iu_0J_0)+f(r)u_1(-J_1\cosh M+iJ_2\sinh M)\big)d\th \nn
  \eea
  with
  \bea
  M&=&\frac{\n^2+3}{4\n}(2t +(\n(r_++r_-)-\sqrt{(\n^2+3)r_+r_-})\th),\nn\\
  f(r)&=&\sqrt{(r-r_+)(r-r_-)},\nn\\
  g(r)&=&r-\frac{\sqrt{(\n^2+3)r_+r_-}}{2\n}.\nn
  \eea
Unlike the pure AdS$_3$ gravity, due to the shortage of the gauge symmetry, the holonomies of the gauge potential do not give the global charges of the black hole.

\section{Null Solution}

For the null warped AdS$_3$, its Killing vectors are given by
\bea
N_1&=&\p_-,\nn\\
N_0&=&x^-\p_-+\frac{u}{2}\p_u, \nn\\
N_{-1}&=&(x^-)^2\p_--u^2\p_++x^- u\p_u,\nn\\
N&=&\p_+.
\eea
One could take $SL(2,R)$ generators $N_{\pm 1},N_0$ as the SKV, then one obtain that
\be
A=A_0dx^++\frac{-A_0(x^+)^2+G_0x^++F_0}{u^2}dx^-+\frac{-2A_0x^++G_0}{u}du,
\ee
with $A_0,G_0,F_0$ being constant matrices. Such kind of gauge potential cannot lead
to null warped AdS$_3$ spacetime. Certainly, it may give us some other configurations.

It turns out that we should choose $U(1) \times U(1)_{N}$ Killing vectors $N_0$ and $N$ as
the SKV's. As a result, we find that the gauge potential should take the following form
\be
A=C_+dx^++\left(\frac{C^u_-}{u^2}+\frac{C^-_-}{x^-}\right)dx^-
+\left(\frac{C^u_u}{u}+\frac{C^-_u}{\sqrt{x^-}}\right)du,
\ee
where $C_+, C^u_-,C^-_-, C^u_u, C^-_u$ are the constant matrices to be determined.
To simplify our discussion, we set $C^-_-, C^-_u$ to be vanishing and denote $C^u_-,C^u_u$ as $C_-,C_u$ respectively, then we get a set of equations from the equations of motion
\bea
[C_+, C_-] &=& [\bar{C}_+, \bar{C}_-], \nn \\
\left[ C_+, C_u\right] &=& [\bar{C}_+, \bar{C}_u], \nn\\
-2C_--[C_-, C_u]&=&-2\bar{C}_--[\bar{C}_-,\bar{C}_u],\nn\\
(1-\frac{1}{\m})[C_+, C_-]&=&\frac{1}{2\m}([B_+,C_-]-[B_-,C_+]), \nn\\
(1-\frac{1}{\m})[C_+, C_u]&=&\frac{1}{2\m}([B_+,C_u]-[B_u,C_+]), \nn\\
(1-\frac{1}{\m})(-2C_--[C_-, C_u])&=&\frac{1}{2\m}(-2B_--[B_-,C_u]+[B_u,C_-]), \nn\\
(1+\frac{1}{\m})[\bar{C}_+, \bar{C}_-]&=&\frac{1}{2\m}([B_+,\bar{C}_-]-[B_-,\bar{C}_+]), \nn\\
(1+\frac{1}{\m})[\bar{C}_+, \bar{C}_u]&=&\frac{1}{2\m}([B_+,\bar{C}_u]-[B_u,\bar{C}_+]), \nn\\
(1+\frac{1}{\m})(
-2\bar{C}_--[\bar{C}_-, \bar{C}_u])&=&\frac{1}{2\m}(-2B_--[B_-,\bar{C}_u]+[B_u,\bar{C}_-]). \nn
\eea

The null solution turns out to be
\bea
A=a_1L_1 dx^++\frac{a_2L_1+a_3L_{-1}}{u^2}dx^-+\frac{a_0L_0}{u}du,\nn\\
\bar{A}=\bar{a}_1L_1 dx^++\frac{\bar{a}_2L_1+\bar{a}_3L_{-1}}{u^2}dx^-+\frac{\bar{a}_0L_0}{u}du,\nn\\
\beta=u_1L_1dx^++\frac{u_2L_1+u_3L_{-1}}{u^2}dx^-+\frac{u_0L_0}{u}du.
\eea
As above, solving the equations of motion and then one finds the corresponding metric to be\footnote{we need to do some coordinate redefinition in this case.}
\be
ds^2=\frac{du^2}{u^2}+\frac{dx^+dx^-}{u^2}\pm \frac{(dx^-)^2}{u^4}.
\ee

\section{Conclusion and Discussion}

We have studied the classical solutions of topologically massive gravity and its higher spin generalization in the first-order formulation. We found the AdS pp-wave solution and its higher spin cousins, by requiring suitable asymptotic behavior of the lagrangian multiplier $\beta$. 
These AdS pp-wave solution can receive the higher spin modification. It would be interesting to study such kind of higher spin modified AdS pp-wave spacetime.

To find the solutions not asymptotic to AdS$_3$, we just made ansatz and tried to solve the equations of motion directly without imposing any boundary condition. We introduced the notion of special Killing vector and apply it to find the solutions.  We managed to rediscover the timelike, spacelike and null warped AdS$_3$ spacetimes, whose SKV were assumed to be $SL(2,R)_R, SL(2,R)_L$ and $U(1)\times U(1)_N$ respectively. It turned out that SKV are powerful enough to fix the ansatz on the gauge potentials and Lagrangian field. The equations of motion are transformed into a set of matrix equations, which could be solved in an algebraic way. It is interesting to see whether this set of matrix equations lead to spacetimes beyond warped AdS$_3$. As SKV's form a subgroup of isometry group, the less strict the SKV, the more difficult to solve the matrix equations.

From our study, we also noticed that usually there are many solutions of gauge potentials, corresponding to exactly one spacetime. One class of degeneracy resides in the gauge potentials, which could be one or even two-parameter class of solutions, corresponding to the same metric. The other class of degeneracy comes from the fact that the different metrics could describe the same spacetime, where the different metrics are from the intrinsic symmetry in the matrix equations. Therefore, this poses an interesting question how to classify the gauge potentials. Notice that though there is no gauge symmetry corresponding to the diffeomorphism, there is gauge symmetry corresponding to the local Lorentz transformation, which relate different gauge potentials to each other.

Moreover, we obtained the spacelike warped AdS black hole through singular coordinate transformation in our framework. Unfortunately, due to the shortage of the gauge symmetry, the holonomy of the gauge potential does not encode the information of the black hole. It seems hard to read the global charges of the black hole from the gauge potentials. Similarly, there are many gauge potentials corresponding to the same black holes.

Another remarkable fact is that for the warped AdS$_3$ spacetime, there exist non-principal embedding as well. In our study, we are free to choose gauge group other than $SL(2,R)$. We showed that in the case of $SL(3,R)$ gauge group, it is possible to consider non-principal embedding, which leads to a spacelike warped AdS$_3$ with different radius. Such phenomenon happens for other warped spacetime.

There are many open questions:
\begin{itemize}
\item How to classify all the solutions in our framework? This question has two-fold meaning. On one side, we need to find a way to classify the gauge potentials in our framework. On the other side, we would like to know if it is possible to find all the solutions of HSTMG.
\item In our study of the warped spacetimes, the final question is how to solve a set of matrix equations. In principle, the equations do not prevent us from considering the gauge groups other than $SL(2,R)$. We have found the AdS pp-wave with nonvanishing higher spin, it would be nice to see if there exists nontrivial higher spin warped spacetime;
\item It would be interesting to discuss the warped black holes with higher spin hair. In our study of warped spacetime, we started from the ansatz constrained by the SKV. This usually gives global warped spacetime. After a singular coordinate transformation, then one arrive at the corresponding black hole solution. Therefore, once we find a global warped spacetime with nonvanishing higher spin fields, it is possible to get the warped black hole with higher spin hair. Another relevant question, is that once we find the black hole solutions with higher spin hair, how to study its physical properties? It seems that we cannot define higher spin charges. 
 This is one of the most fundamental questions hinders us in higher spin topologically massive gravity. Though at the linearized level we have shown that HSTMG makes perfect sense\cite{Chen:2011yx}, we do not know how to deal with it at the non-linearized level besides constructing so many classical solutions.
\item We did not discuss the higher spin perturbations around the warped spacetime. It would be interesting to study this issue.
    \end{itemize}

\section*{Acknowledgments}

The work was in part supported by NSFC Grant No. 10975005. BC would like to
 thank the organizer and participants of the advanced workshop ``Dark Energy
 and Fundamental Theory" supported by the Special Fund for Theoretical Physics
 from the National Natural Science Foundations of China with grant no.: 10947203
 for stimulating discussions and comments.

\section*{Appendix A: Conventions}
In this appendix we specify our convention through out this paper. The $SL(N,R)$ algebra and generators can be found in the paper \cite{Castro:2011iw}. In our paper, when we consider the spin 2 case, we always use
 $L_i=W^2_i$ for simplicity. We also use $J_0,J_1,J_2$, these $J_0,J_1,J_2$ are related to $L_0,L_1,L_{-1}$ by
\be
J_0=\frac{1}{2}(L_1+L_{-1}), J_1=\frac{1}{2}(L_1-L_{-1}),J_2=L_0.
\ee
When we consider the spin 3 case, we use the notation $W_m=W^3_m$ for simplicity.


\section*{Appendix B: Other solutions corresponding to spacelike warped AdS$_3$}

Similar to the timelike warped case, we may use the symmetry to get the other solutions. It turns out that we can set $\ta_0=i\ta_1, \bar{\ta}_0=i\bar{\ta}_1, \tu_0=i\tu_1$ to find the solution of equations of motion. In this case, we have
\bea
\ta_0=-\frac{\mu-3}{\sqrt{\mu^2+27}}, & & \ta_2=-i\frac{(\mu-3)(\mu+9)}{\mu^2+27},\nn\\
\bar{\ta}_0=-\frac{\mu+3}{\sqrt{\mu^2+27}}, & &\bar{\ta}_2=-i\frac{(\mu+3)(\mu-9)}{\mu^2+27},\nn\\
\tu_0=-\frac{2(\mu-3)(\mu+3)}{3\sqrt{\mu^2+27}},&&\tu_2=-i\frac{8(\mu-3)\mu(\mu+3)}{3(\mu^2+27)}
\eea
and
\bea
A&=&(i\ta_0 J_0)d\phi+(-\ta_0J_1\cosh\phi-\ta_2J_2\sinh\phi)d\r\nn\\
&&+(i\ta_0J_0\sinh\r+\ta_0J_1\sinh\phi\cosh\r+\ta_2J_2\cosh\phi\cosh\r)d\t \label{Aw1}
\eea
and similar expression for $\bar{A},\b$.
The metric of the solution is of the form
\bea
ds^2&=&\frac{1}{v^2+3}\left\{(d\phi+\sinh\r d\t)^2+(\cosh\phi d\r-\sinh\phi\cosh\r d\t)^2\right. \nn\\
& &\left. +\frac{4v^2}{v^2+3}(\sinh\phi d\r-\cosh\phi\cosh\r d\t)^2\right\}.\label{metricw1}
\eea

On the other hand, we may set $\ta_0=i\ta_2, \bar{\ta}_0=i\bar{\ta}_2, \tu_0=i\tu_2$ as well, which leads to
\bea
A&=&(i\ta_0 J_0)d\phi+(i\ta_0J_2\sinh\phi-i\ta_1J_1\cosh\phi)d\r \nn\\
& &+(i\ta_0J_0\sinh\r+i\ta_1J_1\sinh\phi\cosh\r-i\ta_0J_2\cosh\phi\cosh\r)d\t \label{Aw2}\\
\bar{A}&=&(i\bar{\ta}_0 J_0)d\phi+(i\bar{\ta}_0J_2\sinh\phi-i\bar{\ta}_1J_1\cosh\phi)d\r\nn\\
& &+(i\bar{\ta}_0J_0\sinh\r+i\bar{\ta}_1J_1\sinh\phi\cosh\r-i\bar{\ta}_0J_2\cosh\phi\cosh\r)d\t
\eea
with
\bea
\ta_0=-\frac{\mu-3}{\sqrt{\mu^2+27}}, & & \ta_1=-i\frac{(\mu-3)(\mu+9)}{\mu^2+27},\nn\\
\bar{\ta}_0=-\frac{\mu+3}{\sqrt{\mu^2+27}}, & &\bar{\ta}_1=-i\frac{(\mu+3)(\mu-9)}{\mu^2+27}.
\eea
The metric of the solution is of the form
\bea
ds^2&=&\frac{1}{v^2+3}\left\{(d\phi+\sinh\r d\t)^2+\frac{4v^2}{v^2+3}(\cosh\phi d\r-\sinh\phi\cosh\r d\t)^2\right. \nn\\
& &\left. -(\sinh\phi d\r-\cosh\phi\cosh\r d\t)^2\right\}.\label{metricw2}
\eea

It is also remarkable the other two warped spacetime satisfy these relations as well. For the spacetime (\ref{metricw1}), we have
\bea
\CT_0=i\ta_0J_0, & & \bar{\CT}_0=i\bar{\ta}_0J_0,\nn\\
\CT_1=\ta_2J_2, & & \bar{\CT}_1=\bar{\ta}_2 J_2, \nn\\
\CT_2=\ta_0J_1, && \bar{\CT}_2=\bar{\ta}_0 J_1
\eea
and for the spacetime (\ref{metricw2}), we have
\bea
\CT_0=i\ta_0J_0, & & \bar{\CT}_0=i\bar{\ta}_0J_0,\nn\\
\CT_1=-i\ta_0J_2, & & \bar{\CT}_1=-i\bar{\ta}_0 J_2, \nn\\
\CT_2=i\ta_1J_1, && \bar{\CT}_2=i\bar{\ta}_1 J_1.
\eea

\end{document}